\documentstyle[mprocl]{article}

\newcommand{\plabel}{\label}

\begin{document}

\renewcommand{\thefootnote}{\fnsymbol{footnote}}

\title{Exact Path Integral Quantization of
2-D Dilaton Gravity 
}


\author{W. Kummer, H. Liebl 
and D.V. Vassilevich
\footnotemark[6]}

\footnotetext[6]{Permanent address: Department of Theoretical Physics,
St. Petersburg University, 198904 St. Petersburg, Russia}

\address{Institut f\"ur
    Theoretische Physik, Technische Universit\"at Wien \\ Wiedner
    Hauptstr.  8--10, A-1040 Wien \\ Austria}


\maketitle
\abstracts{
We demonstrate that in the absence of `matter' fields to all orders of 
perturbation theory and for all 2D dilaton theories the quantum effective 
action coincides with the classical one.  
This resolves the apparent contradiction between the well established
results of Dirac quantization and perturbative (path-integral) 
approaches which seemed to yield non-trivial quantum corrections.  
For the Jackiw--Teitelboim (JT) model, our result is even extended
to the situation when a matter field is present. }

In recent years, stimulated by the `dilaton black 
hole' \cite{dilat} numerous studies of quantized gravity 
in $d=2$ were performed \cite{KLVcomment}
using the second order formalism
\begin{equation}
 \plabel{lbegin}
{\cal{L}}_{(1)}=\sqrt{-g}
\left(-X\frac{R}{2}-\frac{U(X)}{2}(\nabla X)^2+V(X))\right) \quad .
\end{equation}  
A common feature of all these studies is that due to the particular structure
of the theory the constraints can be solved exactly, yielding a finite
dimensional phase space. 
This remarkable property raises hope that one will be able to get insight 
into the information paradox. However, in the presence of an additional 
matter field again an infinite number of modes must  be quantized. 

Using the path integral approach one--loop quantum corrections 
have been considered perturbatively \cite{odintsov} yielding
even for pure dilatonic gravity (\ref{lbegin})  a highly 
non--trivial renormalization structure, clearly in contradiction 
to the results from the above approaches.

Here we close the gap between these two
approaches by demonstrating that for (\ref{lbegin}) there are no local quantum
corrections for the path integral approach as well.
Adding matter fields we are still able to quantize the JT-model \cite{jackiw}
exactly. Finally we comment on higher loop contributions for the general case.

\section{Exact Path Integral Quantization -- Matterless Case}
\label{chap62}

Let our starting point be the first order action ($X^+$, $X^-$ and $X$
are auxiliary fields)
\begin{equation}
  \plabel{lfirst}
  {\cal{L}}_{(2)}=X^+De^-+ X^-De^++Xd\omega +\epsilon(V(X)+X^+X^-U(X)) ,
\end{equation}
where $De^a$, $d\omega$ and $\epsilon$ are the torsion, curvature
and volume two form, respectively.
The quantum equivalence of (\ref{lfirst}) to the second order form (\ref{lbegin})
was demonstrated in \cite{KLV}.
We will be working in a `temporal' gauge $e_0^+ = \omega_0=0, \quad e_0^-=1$
which, by applying the canonical BVF\cite{brs} formalism, produces the 
determinant\cite{KLV}
$F=(\det \partial_0 )^2 \det (\partial_0 + X^+U(X))$.
The generating functional for the Green functions is
\begin{equation}
  \plabel{gen}
  W=\int ({\cal D}X)({\cal D}X^+)({\cal D}X^-)({\cal D}e^+_1)({\cal D}e^-_1)({\cal D}\omega_1)F\exp
\left[i\int_x {\cal{L}}_{(2)} +{\cal{L}}_s \right] ,
\end{equation}
where ${\cal L}_s$ denotes the contribution of the sources $(j^{\pm},j,J^{\pm},J)$
corresponding to the fields $(e^{\mp}_1,\omega_1,X^{\mp},X)$.
Integrating over the zweibein components and the spin connection results in
\begin{eqnarray}
  \plabel{gen2}
  W&=&\int ({\cal D}X)({\cal D}X^+)({\cal D}X^-) \delta_{(1)}\delta_{(2)}
\delta_{(3)}F \exp \left[i \int J^+X^- +J^-X^+ +JX \right] \\
  \plabel{delta}
  \delta_{(1)}&=&\delta \left(-\partial_0 X^+ +j^+ \right) \\
 \delta_{(2)}&=&\delta \left( -(X^+U(X)+\partial_0) X^- +j^- -V(X) \right) \\
 \delta_{(3)}&=&\delta \left(-\partial_0 X +X^+ +j \right) .
\end{eqnarray}
We must stress that due to the partial integrations where 
we discarded the surface terms the following results are true 
only locally. A properly regularized definition of the 
Green functions like $\partial_0^{-1}$, $\partial_0^{-2}$ is presented
in \cite{KLV2}. The remaining integrations can be performed most conveniently 
in the order $X^+, X^-$ and $X$ to yield
\begin{equation}
  \plabel{Z}
  Z=-i \ln W =\int JX +J^-\frac{1}{\partial_0}j^+ +
J^+ \frac{1}{\partial_0 +U(X)\frac{1}{\partial_0}j^+}
\left(j^- -V(X)\right),
\end{equation}
where $X$ has to be replaced by $X=\partial_0^{-2}j^+ +\partial_0^{-1}j$.
Note that the determinant $F$ is precisely canceled by these last three 
integrations. Eq.(\ref{Z}) gives the exact non-perturbative generating functional
for connected Green functions and it does not contain any 
divergences, because it clearly describes tree--graphs only. 
Hence no quantum effects remain.

In \cite{KLV} it was demonstrated that by
inserting the explicit expressions for the mean fields 
(e.g. ${\bar X^\pm} = \frac {\delta Z}{\delta J^\mp}$) into 
the effective action $\Gamma$
one obtains exactly the classical action (\ref{lfirst})
in our temporal gauge, up to surface terms. As expected this clearly 
re-demonstrates
the absence of (local) quantum effects.

\section{Exact Path Integral Quantization -- JT-Model with Matter}
\label{chap63}

Coupling a scalar field minimally \footnotemark[3] 
\footnotetext[3]{The case of non-minimal coupling has been treated in \cite{KLV3}}
to the gravitational action leads to the well
known {\it Polyakov} action
${\cal L}_P=\sqrt{-g}R \Box^{-1} R$.
Clearly ${\cal L}_P$ is not linear in the zweibein 
anymore and will therefore, in general, 
destroy the procedure of section \ref{chap62}. However, for the
JT model \cite{jackiw}
($U(X)=0$, $V(X)=\mbox{const}X$),
we are able to perform a complete integration of the gravitational
{\bf and} the matter action. 
Beginning with the $X$, $X^+$
and $X^-$ integration we arrive (starting from (\ref{gen})) at 
\begin{eqnarray}
 W&=&\int ({\cal D}e^+_1)({\cal D}e^-_1)({\cal D}\omega_1)
\delta_{(X)}\delta_{(X^+)}
\delta_{(X^-)}F \exp i \int j\omega_1 + j^+ e^-_1 + j^- e^+_1 +{\cal L}_P \\
  \plabel{deltaX}
  \delta_{(X)}&=&\delta \left(\partial_0 \omega_1 +J-\alpha e^+_1 \right) \\
  \plabel{deltaX2}
 \delta_{(X^+)}&=&\delta \left(\omega_1 +J^- +\partial_0 e^-_1 \right) \\
\plabel{deltaX3}
 \delta_{(X^-)}&=&\delta \left(\partial_0 e^+_1  +J^+ \right) \quad .
\end{eqnarray}
The remaining integrations can be done with the use of
(\ref{deltaX} to \ref{deltaX3}) during which the term $F$ gets cancelled again.
As a final result we arrive at 
$Z=-i \ln W = \int j\omega_1 + j^+ e^-_1 + j^- e^+_1 +{\cal L}_P $  
where the zweibein and connection have to be expressed as
the solutions of (\ref{deltaX} to \ref{deltaX3}).
This gives the exact non-perturbative
generating functional for connected Green functions,
even in the presence of matter.
In the absence of external matter sources therefore the quantum
JT model with matter is locally equivalent to the classical JT
model with Polyakov term, i.e. the `semiclassical' approximation
becomes exact.

\section{Global Considerations and the General Case with Matter}
\plabel{chap64}

All Dirac quantization approaches treat
models whose kinetic dilaton was removed by a conformal transformation,
which drastically change the global
structure already at the classical level \cite{kat95}.
It is therefore by no means clear
how the quantum theory is affected. As we demonstrated here, {\it local}
quantum effects are not affected by that transformations, however
there are sources of conformal non--invariance
\cite{ccmann} which may change global effects. 

Two-loop contributions from scalar matter for $U(X)=0$ and arbitrary $V(X)$
have been considered in \cite{KLV2}. To this order, interestingly
enough the effective action only experiences a renormalization of the dilaton
potential, i.e. 
\begin{equation}
\Gamma=S_{cl}(\overline{e_1^\pm},\overline{\omega},\overline{X})+
\hbar S_P(\overline{e^-_1},\overline{e^+_1}) +
O(\hbar^3)
\end{equation}
where $V \to V -\hbar^2 \gamma V''$, $\gamma$ being a field independent constant.

\section*{Acknowledgement}

This work has been supported by Fonds zur F\"orderung der
wissenschaftlichen For\-schung (FWF) Project No.\ P 10221--PHY.

\end{document}